\documentclass[sigconf,screen,nonacm]{acmart}

\AtBeginDocument{%
  }

\usepackage[english]{babel}
\usepackage[utf8]{inputenc}
\usepackage{fontawesome}
\usepackage[normalem]{ulem}

\usepackage{xspace}

\graphicspath{{figures/}}
\usepackage{graphicx}
\usepackage{tabularx}
\usepackage{colortbl}
\usepackage{framed}
\usepackage{xcolor}
\usepackage{pgfgantt}
\usepackage{geometry}
\usepackage{tikz}
\usepackage[framemethod=TikZ]{mdframed}
\usetikzlibrary{arrows}
\usetikzlibrary{shapes}

\definecolor{gray}{rgb}{0.5,0.5,0.5}
\definecolor{mauve}{rgb}{0.57, 0.37, 0.43}
\definecolor{rltred}{rgb}{0.5,0,0}
\definecolor{rltgreen}{rgb}{0,0.5,0}
\definecolor{rltblue}{rgb}{0,0,0.5}
\definecolor{webgreen}{rgb}{0, 0.5, 0} %
\definecolor{webblue}{rgb}{0, 0, 0.5} %
\definecolor{webred}{rgb}{0.5, 0, 0} %
\definecolor{weborange}{rgb}{1, 0.6, 0} %
\definecolor{light-gray}{HTML}{B3B3B3}
\definecolor{dim-gray}{HTML}{696969}
\definecolor{white-close}{HTML}{E2E2E2}
\definecolor{light-red}{rgb}{0.968627451,0.8078431373,0.8039215686} %
\definecolor{light-green}{rgb}{0.8392156863,0.9921568627,0.8156862745} %
\definecolor{light-gray}{rgb}{0.8745098039,0.8745098039,0.8745098039} %

\usepackage{array}
\usepackage{multirow}
\usepackage{longtable}
\usepackage{booktabs}

\usepackage{soul} %

\usepackage{amsmath}
\usepackage{amsfonts}
\usepackage{bm}
\usepackage{upgreek}
\usepackage{xfrac}

\usepackage{paralist}
\usepackage{verbatim}
\usepackage{listings}
\usepackage{xpatch}
\makeatletter
\xpatchcmd\l@lstlisting{1.5em}{0em}{}{}
\makeatother

\lstdefinelanguage{Dafny}{
    keywords={method, function, predicate, class, constructor, trait, extends, returns, requires, ensures, modifies, reads, invariant, forall, exists, if, else, true, false, while, for, match, case, var, const, int, nat, real, bool, bv4, char, string, array, set, multiset, seq, map, new, print},
    keywordstyle=\color{blue}, %
    identifierstyle=\color{black},
    comment=[l]{//},
    commentstyle=\color{gray}\ttfamily,
    stringstyle=\color{red}\ttfamily,
    sensitive=true
}

\lstdefinestyle{diff}{
    language=Dafny,
    moredelim=**[is][\lstset{keywordstyle=\color{red}}\color{red}]{--}{--}, %
    moredelim=**[is][\lstset{keywordstyle=\color{green!50!black}}\color{green!50!black}]{++}{++}, %
}

\lstdefinestyle{sharpc}{
    language=[Sharp]C,
    moredelim=**[is][\color{green!50!black}]{++}{++} %
}

\usepackage[inline]{enumitem}
\setitemize{noitemsep,topsep=1pt,parsep=1pt,partopsep=1pt}

\usepackage{caption,subcaption}
\usepackage{float}

\usepackage[hyphens]{xurl}
\hypersetup{
  pdftitle={},
  pdfauthor={},
  colorlinks=true,
  linktoc=all,
  bookmarks=false,
  bookmarksopen,
  bookmarksnumbered
}

\usepackage[hang,flushmargin,multiple]{footmisc}

\usepackage{acronym}
\acrodef{AMR}{Accessor Method Replacement}
\acrodef{AOD}{Arithmetic Operator Deletion}
\acrodef{AOI}{Arithmetic Operator Insertion}
\acrodef{AOR}{Arithmetic Operator Replacement}
\acrodef{AST}{Abstract Syntax Tree}
\acrodef{BBR}{Boolean-Binary Expression Replacement}
\acrodef{BOR}{Binary Operator Replacement}
\acrodef{CBE}{Conditional Block Extraction}
\acrodef{CBR}{Case Block Replacement}
\acrodef{CIR}{Collection Initialization Replacement}
\acrodef{COD}{Conditional Operator Deletion}
\acrodef{COI}{Conditional Operator Insertion}
\acrodef{COR}{Conditional Operator Replacement}
\acrodef{DCR}{Datatype Constructor Replacement}
\acrodef{DSL}{Domain Specific Language}
\acrodef{EVR}{Expression Value Replacement}
\acrodef{FAR}{Field Access Replacement}
\acrodef{GUI}{Graphical User Interface}
\acrodef{JVM}{Java Virtual Machine}
\acrodef{LBI}{Loop Break Insertion}
\acrodef{LHS}{Left-Hand Side}
\acrodef{LINQ}{Language Integrated Query}
\acrodef{LLM}{Large Language Model}
\acrodef{LOD}{Logical Operator Deletion}
\acrodef{LOI}{Logical Operator Insertion}
\acrodef{LOR}{Logical Operator Replacement}
\acrodef{LSR}{Loop Statement Replacement}
\acrodef{LVR}{Literal Value Replacement}
\acrodef{MAP}{Method Argument Propagation}
\acrodef{MCR}{Method Call Replacement}
\acrodef{MMR}{Modifier Method Replacement}
\acrodef{MNR}{Method Naked Receiver}
\acrodef{MRR}{Method Return Value Replacement}
\acrodef{MSG}{Mutant Schemata Generation}
\acrodef{MVR}{Method-Variable Replacement}
\acrodef{ODL}{Operator Deletion}
\acrodef{OO}{Object-Oriented}
\acrodef{PRV}{Polymorphic Reference Replacement}
\acrodef{RHS}{Right-Hand Side}
\acrodef{ROR}{Relational Operator Replacement}
\acrodef{SAR}{Argument Swapping}
\acrodef{SDL}{Statement Deletion}
\acrodef{SLD}{Subsequence Limit Deletion}
\acrodef{SOR}{Shift Operator Replacement}
\acrodef{STP}{Spec-Testing Proof}
\acrodef{SWS}{Statement Swapping}
\acrodef{SWV}{Variable Declaration Swapping}
\acrodef{TAR}{Tuple Access Replacement}
\acrodef{THI}{This Keyword Insertion}
\acrodef{THD}{This Keyword Deletion}
\acrodef{UOD}{Unary Operator Deletion}
\acrodef{UOI}{Unary Operator Insertion}  
\acrodef{VER}{Variable Expression Replacement}
\acrodef{VDL}{Variable Deletion}

\usepackage[capitalise,noabbrev]{cleveref} %

\newcommand{\ourTool}{MutDafny\xspace}

\usepackage[framemethod=TikZ]{mdframed}

\begin{document}

\title{MutDafny: A Mutation-Based Approach to Assess Dafny Specifications}

\author{Isabel Amaral}
\orcid{0009-0009-4188-9233}
\affiliation{
  \institution{INESC TEC, Faculdade de Engenharia, Universidade do Porto}
  \city{Porto}
  \country{Portugal}
}
\email{isabel.andre.amaral@gmail.com}

\author{Alexandra Mendes}
\orcid{0000-0001-8060-5920}
\affiliation{
  \institution{INESC TEC, Faculdade de Engenharia, Universidade do Porto}
  \city{Porto}
  \country{Portugal}
}
\email{alexandra@archimendes.com}

\author{José Campos}
\orcid{0000-0001-7565-8382}
\affiliation{
  \institution{Faculdade de Engenharia, Universidade do Porto \& LASIGE, Universidade de Lisboa}
  \city{Porto}
  \country{Portugal}
}
\email{jcmc@fe.up.pt}

\begin{abstract}

In verification-aware languages, such as Dafny, despite their critical role, specifications are as prone to error as implementations.  Flaws in specifications can result in formally verified programs that deviate from the intended behavior. In this paper, we explore the use of mutation testing to reveal weaknesses in formal specifications written in Dafny. 

We present \ourTool, a tool that increases the reliability of Dafny specifications by automatically signaling potential weaknesses. Using a mutation testing approach, we introduce faults (\emph{mutations}) into the code and rely on formal specifications for detecting them. If a program with a mutant verifies, this may indicate a weakness in the specification. We extensively analyze mutation operators from popular tools, identifying the ones applicable to Dafny. In addition, we synthesize new operators tailored for the language from bugfix commits in publicly available Dafny projects on GitHub. Drawing from both, we equipped our tool with a total of 40 mutation operators. 
We evaluate \ourTool's effectiveness and efficiency on a dataset of 794 real-world Dafny programs, and manually analyze a subset of the resulting undetected mutants, identifying five weak real-world specifications (on average, one at every 241 lines of code) that would benefit from strengthening. 

\end{abstract}

\maketitle

\section{Introduction}\label{chap:intro}

Proving that a software program is correct is necessary in critical systems, as failures in these can have drastic consequences. It is thus important to guarantee that such systems have no faults. Although traditional software testing~\cite{Myers2011} is used in many contexts for software validation, it only checks a system against specific inputs, not guaranteeing full correctness. In contrast, \emph{formal software verification}~\cite{leroy2009formally, klein2009sel4, beringer2015verified} can mathematically prove complete correctness concerning a precise specification of the intended behavior. 

Verification-aware languages, such as Dafny~\cite{dafny}, are designed to support the development of formally verified software by embedding formal specifications directly into the code. It allows developers to specify program behavior through pre-conditions (\emph{requires}), post-conditions (\emph{ensures}), and invariants (\emph{invariant}), which are automatically verified by Dafny, helping to guarantee that the implementation conforms to its intended behavior and ensuring code correctness beyond conventional testing. Dafny has seen adoption in industry, including at AWS\footnote{\url{https://github.com/aws/aws-encryption-sdk-dafny}, accessed Jan. 2026.}~\cite{chakarov2025formally} and Consensys~\cite{cassez2021verification}. 

Since formal verification only guarantees that an implementation conforms to its specification, a flawed specification can lead to a verified program that does not exhibit the intended behavior. Although specifications are usually trusted to be correct, this should not be taken for granted\footnote{In this paper, we use \emph{specification} or \emph{spec} to refer to Dafny's formal contracts (e.g., requires, ensures, and invariants); we use \emph{implementation} or \emph{code} to refer to Dafny's code within method or function bodies.}.
A study conducted by \citet{Fonseca2017} confirmed exactly that: it found 16 faults in three formally verified distributed systems (one written in Dafny). Some of these faults stemmed from incorrect and incomplete specifications, highlighting a critical limitation of verification when specifications themselves are flawed.

As a motivational example, consider Dafny's code in \Cref{lst:weak-spec-correct-impl} returning all elements belonging to both input arrays.

\begin{lstlisting}[style=diff, label=lst:weak-spec-correct-impl, caption={Example of a Dafny program with a weak specification.}, escapechar=|]
predicate InArray(a: array<int>, x: int)
reads a
{ exists i :: 0 <= i < a.Length && a[i] == x }

method SharedElements(a: array<int>, b: array<int>) 
    returns (result: seq<int>)
ensures forall x :: x in result ==> (InArray(a, x) && InArray(b, x))|\label{line:firstEnsure}|
ensures forall i, j :: 0 <= i < j < |result| ==> result[i] != result[j]
{
     var res: seq<int> := [];
     for i := 0 to a.Length|\label{line:betweenA}|
     invariant 0 <= i <= a.Length
     invariant forall x :: x in res ==> InArray(a, x) && InArray(b, x)
     invariant forall i, j :: 0 <= i < j < |res| ==> res[i] != res[j]
     {|\label{line:betweenB}|
---        if InArray(b, a[i]) && a[i] !in res { --|\label{line:beginIf}|
---            res := res + [a[i]]; --|\label{line:bodyIf}|
---        } --|\label{line:endIf}|
     }
     result := res;
}
\end{lstlisting}

Although this program verifies, removing the \texttt{if} block in Lines \ref{line:beginIf} to \ref{line:endIf} (an error a human could make or a mutation tool could insert)
causes it to always return an empty list\,---\,an incorrect behavior that still satisfies the specification. Despite the inserted error, Dafny's ability to verify this program hints at a possible problem with the spec. This is indeed an example of a \emph{weak} specification, i.e., one that does not sufficiently constrain the program's behavior. Its strengthening would require replacing the \texttt{==>} in the first \texttt{ensures} (Line \ref{line:firstEnsure}) with \texttt{<==>} and the addition of an invariant (between Lines \ref{line:betweenA} and \ref{line:betweenB}) adding that, besides every element in \texttt{result} being in both arrays, every element that appears in both arrays must also be present in \texttt{result}, i.e.:
\begin{lstlisting}[style=diff, numbers=none, label=lst:strong-spec-correct-impl]
+++ invariant forall j :: 0 <= j < i && InArray(b, a[j]) ==> a[j] in res ++
\end{lstlisting} 

In this work, our main goal is to support Dafny developers in ensuring the strength of their specifications. To achieve this, we follow the \emph{mutation testing}~\cite{Woodward1993,Yue2011,Sanchez2024,Gopinath2022} concept, where the quality of a test suite is evaluated by introducing faults, called \emph{mutants}, into a program, and evaluating whether these are detected by the test suite.

Mutants are automatically created by applying \emph{mutation operators}, i.e., syntactic changes that mimic common programming errors, each differing from the original program by a small localized change. Mutants that cause tests to fail are considered \emph{killed}, while those that go undetected are \emph{alive}, indicating untested behavior. While the mutants typically run against a test suite, in our approach, they run against the formal specification. Surviving mutants may reveal weaknesses in the spec, indicating underspecified behavior, which may help developers identify and strengthen those specs. Nevertheless, an alive mutant may not always indicate the presence of a weak specification; i.e., the mutant might just be equivalent to the original program~\cite{Budd1982,Yao2014,Madeyski2014}, meaning that, despite the change, its behavior remains the same as that of the original program. 

This raises the questions: \textbf{(RQ1)} \emph{Which small syntactic changes could mimic programming errors in Dafny?} and \textbf{(RQ2)} \emph{Are mutation operators for Dafny programs helpful in the identification of specification weaknesses?} To answer these, we first assembled a compilation of 147 mutation operators previously proposed for other six programming languages and studied their application in Dafny. Secondly, we complemented this set with new operators tailored for the language, derived from an analysis of 1,475 bugfix commits from Dafny programs available in 112 public GitHub repositories. Thirdly, since, to our knowledge, there is no existing tool that mutates Dafny implementations, we developed one, \textbf{\ourTool}, supporting these mutation operators. Finally, we conducted an empirical study on 118,458 pairs of original-mutant Dafny programs to assess the effectiveness of our approach and, additionally, answer the questions: \textbf{(RQ3)} \emph{How effective are the different mutation operators?} and \textbf{(RQ4)} \emph{How efficient is MutDafny at generating mutants?}

\smallskip\noindent
The main \textbf{contributions} of this paper are:
\begin{enumerate}[leftmargin=*]
  \item[\small{$\bigstar$}] An extensive analysis of which mutation operators supported by the most popular mutation tools can be applied to Dafny programs, and why.
   (\Cref{chap:literature-mutation-operators})

  \item[\small{$\bigstar$}] The proposal of a set of novel mutation operators tailored for Dafny programs. (\Cref{chap:dafny-mutation-operators})

  \item[\small{$\bigstar$}] \ourTool: a novel mutation testing tool for signaling possible specification weaknesses in Dafny programs. The tool is publicly available at \url{https://github.com/MutDafny/mutdafny}. (\Cref{chap:tool})

  \item[\small{$\bigstar$}] Evidence of \ourTool's usefulness in assisting Dafny developers increase the reliability of their specs, gathered through an empirical study on 794 Dafny programs and a manual evaluation of 284 alive mutants. All the scripts, data, and instructions required to reproduce the study are publicly available at \url{https://github.com/MutDafny/mutdafny-study-data}. (\Cref{chap:eval})

\end{enumerate}

Even though this paper explores mutation testing in the context of Dafny, an added benefit is that this approach is applicable to any verification-aware language.

\section{Related Work}\label{chap:rw}

\smallskip\textbf{Mutations in verification-aware languages.}
To our knowledge, IronSpec\footnote{\url{https://github.com/GLaDOS-Michigan/IronSpec}, accessed Jan. 2026.}~\cite{IronSpec} is the only previous work dealing with mutations in Dafny, being the work more closely related to ours. It aims to increase the reliability of formal specifications by operating on three different dimensions, one involving mutation testing, but unlike our approach, mutating specifications. Their technique considers mutants that are stronger than the original specification. If the method still verifies with the mutation, it means that the original specification is weaker than the implementation and can potentially be strengthened. In 14 Dafny methods, IronSpec detected two specification faults through this approach.

Much like in \ourTool, mutations to the implementation of Eiffel programs have been used to evaluate their specs~\cite{eiffel}. However, no developer-oriented tool based on the mutation of implementations exists for providing specification quality feedback, nor has any in-depth study been conducted on suitable mutation operators.

Other related work in using mutation testing to assess specification quality includes \citet{spec-generation-eval1} and \citet{Endres2024}. \citet{spec-generation-eval1} uses the mutation of test cases to validate specifications without the need for an implementation. \citet{Endres2024} capture a specification's ability to reject LLM-generated implementation mutants. %

\smallskip\noindent\textbf{{Mutation in other specification-related contexts.}}
MuAlloy~\cite{mualloy} is a mutation testing and mutation-based test generation tool for Alloy. Although both Alloy and Dafny support formal methods, the former is a specification modeling language with a distinct purpose from verification-aware ones.
\citet{reactive-synthesis} used mutations consisting of the deletion of specification guarantees (weakening) to validate a coverage-based scenario generation algorithm for systems generated by reactive synthesis. Finally, in a mutation-based study on the effectiveness of the JML-JUnit~\cite{jml-junit} tool for automatic unit test generation from Java specifications, \citet{jml-junit-study} concluded that mutation testing can also help unveil specification bugs, not just implementation errors.

\smallskip\noindent\textbf{Synthesis of mutation operators.}
Unlike our manual approach, \citet{8919234} used deep learning trained on 787,178 bug-fixing commits from GitHub to automatically generate mutants closely resembling real faults. However, replicating their approach is difficult due to the need for large-scale bugfix data and the fact that we only identified 1,475 (\emph{likely}) bug-fixing commits (0.19\% of their dataset).

\section{Mutation tools and operators proposed for traditional programming languages}\label{chap:literature-mutation-operators}

Aiming to answer \textbf{RQ1}, in this section, we compile the mutation operators proposed for other, more traditional, programming languages and investigate their applicability to Dafny programs. We consider a mutation operator applicable to a Dafny program if (a) the Dafny programming language supports the operation performed by the operator, and (b) if it has the potential of leading to a valid program.

\subsection{Mutation tools proposed for other programming languages but Dafny}\label{sec:classical-mutation-tools}

\citet{Sanchez2022} conducted a study of the use of mutation testing in practice and identified 127 mutation testing tools. The most popular tools for the programming languages for which Dafny compiles to (C\#, C++, Java, JavaScript, Go, and Python) are: Stryker.NET~\cite{stryker.net} for C\#; Mull~\cite{mull, mull-url} for C++; Major~\cite{major2011, major2014, major}, MuJava~\cite{muJava2005, muJava2006, muJavaMethodLevelOperators, muJavaClassLevelOperators, mujava}, and PIT~\cite{pitest, pit} for Java; StrykerJS~\cite{strykerjs} for JavaScript; Go-Mutesting~\cite{go-mutesting1, go-mutesting2} for Go; Mutmut~\cite{mutmut} and MutPy~\cite{mutpy} for Python. These were selected based on the overall top 10 most popular tools and on the most popular for each programming language.

\subsection{Mutation operators proposed for other programming languages but Dafny}\label{sec:classical-mutation-operators}

We identified the set of mutation operators supported by the tools listed in \Cref{sec:classical-mutation-tools}. This involved reading the papers (when available), documentation, and, for Mutmut~\cite{mutmut}, studying its source code due to a lack of documentation.

We found many similar operators across the different tools and languages, but most use different names. We grouped the operators according to our perceived similarity between them: 
\emph{Operator replacement}, which apply to binary operators: arithmetic (e.g., $+$, $*$), relational (e.g., $==$, $<=$), conditional (e.g., $\&\&$, $||$), logical (e.g., $\&$, $|$), shift (e.g., $<<$, $>>>$), assignment (e.g., $+=$, $>>>=$);
\emph{Operator insertion/deletion}, which apply to unary operators: arithmetic (e.g., $-$, $++$), conditional (e.g., $!$), logical (e.g., \textasciitilde);
\emph{Literal value replacement};
\emph{Expression replacement};
\emph{Statement, variable, constant, or operator deletion}\,---\,where the latter differs from unary operator deletion in that it applies to all types of operators, unary and binary, and involves the deletion of all occurrences of a specific operator;
\emph{\ac{OO} specific}: Inheritance, Polymorphism, and Overloading. The operators listed for the arithmetic, relational, conditional, logical, shift, and assignment groups are those specific to Java.

\subsection{Mutation operators proposed for other languages that are applicable in Dafny}

Following the process detailed in \Cref{sec:classical-mutation-operators}, we compiled a total of 147 mutation operators\footnote{The list of mutation operators, including their description, which tools implement them, and their applicability to Dafny, can be found in the \emph{supplementary material} at \url{https://doi.org/10.6084/m9.figshare.30640202.v2}.}. Of these, we identified 97 that can be applied to Dafny programs and 50 that cannot due to the syntax structure of Dafny 4.10.0.\footnote{At the time of writing, the latest version of Dafny available was 4.10.0.} From the 97 operators that can be applied to Dafny, we concluded that seven always generate equivalent mutants when applied to Dafny. Finally, the 90 operators were narrowed down to 30 unique ones, as many reflected the same behavior. Next, we partially answer \textbf{RQ1} by listing and describing the applicability of each operator and any usage restrictions.

\subsubsection{Arithmetic operators}

Dafny provides five arithmetic binary ($+$, $-$, $*$, $/$, $\%$) and a single unary operator ($-$). Shortcut arithmetic operators ($++$, $--$), although common in other programming languages, are not available in Dafny.

\smallskip\emph{\ac{AOR}}~\cite{major2011, major2014, muJavaMethodLevelOperators, pitest, stryker.net, strykerjs, mutpy, mutmut, mull, go-mutesting2}
replaces an arithmetic binary operator by another. However, Dafny does not allow the out-of-the-box replacement of an arithmetic operator with $/$ or $\%$, as the verifier requires a pre-condition stating that the divisor is not zero.
One approach adopted by some tools consists of restricting which operators can be replaced by which. In Dafny, this reflects in allowing free replacement between the $+$, $-$, and $*$ group of operators, and only allowing $/$ to be replaced with $\%$ and vice-versa. Replacement of unary arithmetic operators cannot be applied in Dafny, as it only supports one unary operator ($-$).

\smallskip\emph{\ac{AOI}}~\cite{muJavaMethodLevelOperators, pitest} and \emph{\ac{AOD}}~\cite{muJavaMethodLevelOperators, pitest, mutpy} can only apply to unary operators, namely, to Dafny's unary minus ($-$).

\subsubsection[Relational Operator Replacement]{\protect\acf{ROR}~\cite{major2011, major2014, muJavaMethodLevelOperators, pitest, mutpy, mull, stryker.net, strykerjs, go-mutesting1, go-mutesting2, mutmut}}

Dafny supports the six relational operators common to most languages ($==$, $!=$, $<$, $<=$, $>$, $>=$) and their replacement can be applied without any restriction.
Additionally, the replacement of expressions using relational operators with \texttt{true} or \texttt{false} is also commonly implemented and possible in Dafny. Henceforth, we denote this as \emph{\ac{BBR}}~\cite{muJavaMethodLevelOperators}.

\subsubsection{Conditional operators}\label{sec-conditional-operators}

Dafny supports three conditional operators: two binary ($\&\&$, $||$) and one unary ($!$).

\smallskip\emph{\ac{COR}}~\cite{major2014, muJavaMethodLevelOperators, mutpy, mull, stryker.net, strykerjs, mutmut, go-mutesting1, go-mutesting2}
replaces a binary conditional operator with an alternative. As in \emph{\ac{ROR}}, many tools replace conditionals with \texttt{true} or \texttt{false}, another use of \emph{\ac{BBR}}~\cite{major2014, pitest, stryker.net, strykerjs, go-mutesting1, go-mutesting2, mutmut}.

\smallskip\emph{\ac{COI}}~\cite{muJavaMethodLevelOperators, stryker.net, strykerjs, mutpy, mutmut} and \sloppy\emph{\ac{COD}}~\cite{muJavaMethodLevelOperators, mutpy, mutmut} can be both applied to Dafny's unary conditional operator (\texttt{!}).

\subsubsection{Logical operators}\label{sec:logicalOperators}
Dafny's binary logical bitwise operators are: $\&$, $|$, and \texttt{\^}, and its single unary operator is $!$. While $!$ can be both conditional and logical, it is used in different contexts, either for boolean expressions or for bit-vector ones, returning different types accordingly.

\smallskip\emph{\ac{LOR}}~\cite{major2011, major2014, muJavaMethodLevelOperators, pitest, mutpy, mutmut, stryker.net, go-mutesting2} can be applied to Dafny's binary logical operators without restrictions, replacing them interchangeably. It cannot be applied to the unary operator as there is only one. 

\smallskip\emph{\ac{LOI}}~\cite{muJavaMethodLevelOperators} and \emph{\ac{LOD}}~\cite{muJavaMethodLevelOperators, mutpy, mutmut} can be applied to Dafny without restrictions, i.e., any bit-vector expression in Dafny can be modified by inserting or removing the $!$ operator.

\subsubsection{\acf{SOR}~\cite{major2011, major2014, muJavaMethodLevelOperators, mutmut, stryker.net, go-mutesting2}}
Like \hyperref[sec:logicalOperators]{\emph{logical operators}}, shift ones operate over bit-vector types. Dafny supports the binary shift operators $<<$ and $>>$. Since there are no unary shift operators, insertion or deletion cannot be applied.

\subsubsection{\acf{LVR}~\cite{major2011, major2014, pitest, stryker.net, strykerjs, mutpy, mutmut, go-mutesting2, mull}}
This operator is traditionally applied to boolean, string, and numerical literals, with modifications that vary significantly between tools. The simplest case is the boolean: the only possible replacement is the logical complement, i.e., \texttt{true} with \texttt{false} and vice-versa. For string literals, common transformations include replacing non-empty with empty ones, default values, or inserting random characters in any position. Numerical literals are often incremented, decremented, negated, or replaced by constants, e.g., $0$, $1$, and $-1$. \emph{\ac{LVR}} can be applied to Dafny without restriction.

\subsubsection{\acf{EVR}~\cite{major, pitest, mutmut}}\label{sec-expression-value-replacement}
Corresponds 
to the replacement of primitive-typed expressions (e.g., numerical, boolean) with literal values. Given the broad definition of what constitutes an expression in a program, mutation tools implement a wide range of expression value replacement operators. Some target any primitive-typed expression throughout the code, while others target specific contexts (e.g., return values) and perform more specific replacements (e.g., \texttt{true}, \texttt{false}, and \texttt{null} returns).

\smallskip\emph{Method call replacement.}
Some tools replace method calls with a default value matching their return type, applying only to non-void methods. In Dafny, methods may return multiple values, meaning the replacement may include several defaults. This distinction from \hyperref[sec-expression-value-replacement]{\emph{\ac{EVR}}}, makes it a distinct operator, \emph{\ac{MRR}}~\cite{pitest, mull}.

A more specific method call replacement operator is the \emph{\ac{MAP}}~\cite{pit}, which replaces a method call with one of its arguments. This must match the method's return type to preserve program validity. 

\smallskip\emph{Object initialization replacement.}
An example is the \emph{\ac{CIR}}~\cite{stryker.net, strykerjs} operator, which handles collection initialization with type-specific default constructors. Stryker.NET implements this for C\# arrays, lists, collections, and dictionaries. Although these collection types have different designations in Dafny, this concept extends to them, i.e., arrays, (multi)sets, sequences, and maps. Additionally, in class instances, initialization through a constructor call can be replaced with \texttt{null}~\cite{pitest}.

\subsubsection{Loop constructs replacement.}\label{sec:loop-constructs}
This operator targets loop-specific constructs, replacing \texttt{break} with \texttt{continue} and vice-versa. Replacing both of these with a \texttt{return} is also valid when the enclosing method is \texttt{void}. We refer to this as the \emph{\ac{LSR}}~\cite{mutpy, mutmut, go-mutesting2}. \emph{\ac{LBI}}~\cite{go-mutesting2}, inserts a \texttt{break} at the loop's entry.

It is possible to apply these operators to Dafny programs, but their effectiveness often depends on the structure of the loop. For instance, certain mutations (e.g., inserting a \texttt{continue} before a counter update) may prevent Dafny from verifying loop termination. These can easily cause verification to fail simply due to the application context.

\subsubsection{\ac{CBR}~\cite{pit}}
This operator replaces the bodies of \texttt{case} blocks in switch statements: the default case is replaced with the body of the first non-default label, and the remaining cases are replaced with the default body. Dafny's analogous construct is the \texttt{match} statement, where the default case is denoted by an underscore (\texttt{\_}).

\subsubsection{\acf{SDL}~\cite{major2014, muJavaMethodLevelOperators, stryker.net, go-mutesting1, go-mutesting2}}
Like \hyperref[sec-expression-value-replacement]{\emph{\ac{EVR}}}, statement deletion operators can be categorized by context. The most general form deletes any statement type (e.g., \texttt{return}, \texttt{break}, assignments, calls, etc.) without targeting a specific category. While applicable to Dafny, such deletions may lead to invalid programs (e.g., as in other languages, if a variable's initialization is removed but later used).

\smallskip\emph{Block deletion.}\label{sec:block-deletion-mutation}
An extreme form of statement deletion is removing entire code blocks. In Dafny, this includes deleting method bodies~\cite{stryker.net, strykerjs}, entire \texttt{if}, \texttt{else if}, and \texttt{else} branches, or \texttt{case} blocks in match statements~\cite{go-mutesting1, go-mutesting2}. Method bodies can only be deleted for \texttt{void} methods, and \texttt{if} branches only when no alternatives (i.e., \texttt{else}) are present.

\smallskip\emph{Method call deletion}~\cite{pit, mull}.
Calls to void methods can be deleted without invalidating the program. Deleting non-void calls is not supported (in existing work or in \ourTool), as their return values are typically used elsewhere in the program (e.g., in assignments or expressions), and removing them would result in a syntactically incorrect program. Such cases are better handled by the \emph{\ac{MRR}} operator.

The \emph{\ac{MNR}}~\cite{pit} operator is a specific type of deletion operator for methods that are members of a class. It deletes the method call while preserving the receiver object. To avoid type errors, the receiver must have the same type as the method's return value. Moreover, the operator can only be applied to non-void methods in Dafny.

\smallskip\emph{Field initialization.}
Finally, there are two operators that initialize class fields with arbitrary values through deletion. 

The first targets fields initialized at the declaration~\cite{pit}. Dafny does not provide specific default values for each type but ensures that variables of certain types, called \emph{auto-initializable}, always hold legal values. While it ensures these values are legal, it does not define fixed defaults for each type, assigning arbitrary values.
This feature does not work for user-defined classes, as they are not auto-initializable, and only works for constant fields, as Dafny allows only constants to be initialized at declaration.

The second, implemented by MuJava~\cite{muJava2005, muJavaClassLevelOperators}, deletes a class's constructor, since Java automatically generates a default constructor that initializes fields with default values. Dafny also supports automatic constructor creation, but these are invoked without parentheses, e.g., \texttt{new Car} instead of \texttt{new Car()}.
Applying this operator to a Dafny program would invalidate it, as constructor calls with parentheses would not be valid. An alternative is to delete the constructor body, allowing fields to be initialized with arbitrary values.

\subsubsection{\acf{VDL}~\cite{muJavaMethodLevelOperators}}\label{sec:VDL} This operator deletes all uses of a variable. If the variable is used in a binary expression, the corresponding operator must also be deleted.

Similarly, Python supports slicing as \texttt{s[start\allowbreak:end\allowbreak:step]}, and MutPy~\cite{mutpy} includes an operator that deletes one of the slice elements. Dafny provides a similar slicing syntax for sequences, supporting \texttt{start} and \texttt{end} but not \texttt{step}. Since this operator removes only part of a slice, unlike \emph{\ac{VDL}}, we refer to it as \emph{\ac{SLD}}.

\subsubsection{\acf{ODL}}

Much like the \hyperref[sec:VDL]{\emph{\ac{VDL}}} operator, \emph{\ac{ODL}} deletes all occurrences of a specific operator (e.g., arithmetic, relational). For binary operators, program validity entails deleting one of the operands as well.

\subsubsection{Object-Oriented operators}
MuJava~\cite{muJava2005, muJavaClassLevelOperators} was the first tool to propose a comprehensive set of mutation operators targeting \ac{OO} features like inheritance, polymorphism, and encapsulation. Some of these can be adapted to Dafny, which also supports \ac{OO} features.

\smallskip\emph{\ac{PRV}}\label{sec:polymorphism-operators}
replaces an assignment to a parent class object with a different child class instance. It applies when the parent has multiple children and at least two variables of different child types are in scope.
This may cause side effects if subclasses implement the same methods or initialize fields differently. In the following example, each subclass sets a different value for \texttt{numSides}, so the mutation leads the program to output an incorrect value.

\begin{lstlisting}[style=diff, numbers=none, label=lst:prv-mutation]
      var shape: Shape;
      var rectangle := new Rectangle(10.0, 20.0);
      var triangle := new Triangle();
 ---    shape := rectangle; -- // numSides = 4
 +++    shape := triangle; ++ // numSides = 3
      print shape.numSides, "\n";
\end{lstlisting}

\smallskip\emph{\acf{THI}} and \emph{\acf{THD}.}
These two operators manipulate the \texttt{this} keyword, one by inserting, and the other by deleting it. These apply only when a class method has a parameter with the same name as a class field. 
Neither can be used on the \ac{LHS} of an assignment because Dafny does not allow assignments to parameters: inserting \texttt{this} would require the \ac{LHS} to originally be a parameter, which is not possible, and deleting it would leave a parameter on the \ac{LHS}, invalidating the program.

\smallskip\emph{\acf{AMR}} and \emph{\acf{MMR}} 
replace one method with another having the same signature. The first targets accessor methods, typically starting with \texttt{get}, which access class fields. The second targets modifier methods, usually starting with \texttt{set}, and which update an object's state.

\subsection{Mutation operators proposed for other languages excluded for Dafny}\label{sec:mutnotapplicalable}

Most operators from other languages that cannot be applied to Dafny target syntax constructs that do not exist in the language, such as shortcut operators, compound assignments (e.g., $+$$=$), switch statements, exception-catching blocks, and method overloading. Others rely on language-specific elements like Python's \texttt{None}, Java's \texttt{BigInteger}, or libraries like C\#'s \ac{LINQ}. Some constructs exist in Dafny, but the resulting mutants are invalid or equivalent due to language constraints. Below, we highlight operators excluded for more complex reasons.

\subsubsection{Operator replacement operators}
Major's \emph{Operator Replacement Unary (ORU)}~\cite{major2011, major2014} cannot be applied to Dafny, as it replaces unary operators that do not necessarily belong to the same group, e.g., Java's $-$ and \textasciitilde, which both take numerical arguments. The unary operators supported in Dafny are $-$, for numeric types, and $!$, for booleans or bit-vectors, and they cannot replace each other due to the different argument type requirements.

\subsubsection{Expression replacement operators}
Stryker.NET's \emph{Initialization}~\cite{stryker.net} operator replaces a parameterized constructor with an empty one. Since Dafny only allows one constructor per class, this operator cannot be applied to it.

StrykerJS's \emph{Method Expression}~\cite{strykerjs} replaces JavaScript strings, arrays, and math methods interchangeably, and Stryker.NET's \emph{String Methods}~\cite{stryker.net} does this for .NET string methods (e.g., \texttt{toUpperCase} with \texttt{toLowerCase}). These do not apply to Dafny, since (i) it provides limited built-in support for its types, and (ii) its strings are not objects.

\subsubsection{Statement deletion operators}

Mutmut's \emph{Argument List Mutation}~\cite{mutmut} deletes an element from a method's argument list. This is valid in Python due to the support of optional arguments with default values. Dafny, however, always requires a fixed number of arguments.

\subsubsection{Object-Oriented operators}
It is important to note that Dafny classes are only allowed to extend traits, which act as abstract superclasses and cannot declare constructors or be instantiated. Methods and fields can either be declared in the trait and used in its subclasses, or at the subclass level.

\smallskip\emph{Inheritance.}
MuJava's inheritance-related operators~\cite{muJava2005, muJavaClassLevelOperators} test three different aspects of this feature in Java: method overriding, hiding variables, and the use of the \texttt{super} keyword, none of which are supported in Dafny.

\smallskip\emph{Polymorphism.} 
MuJava's polymorphism operators~\cite{muJava2005, muJavaClassLevelOperators} replace child types with parent types (and vice-versa) in various contexts, such as variable and parameter declarations and type casting. While some are possible in Dafny, the resulting mutants are always equivalent to the original.

Since superclasses cannot be instantiated, object creation always uses subclass constructors. Thus, even when declared with a type of a superclass, Dafny knows the type of the object, using the correct values for its fields and the correct child-specific implementation for methods that are declared but not implemented in the parent.

In mutations that replace child with parent types, if child-specific methods or fields are used, it leads to an invalid program. Otherwise, the mutant is equivalent. The inverse, parent-to-child replacements via type casting, requires prior type checking, e.g., inside of an \texttt{if var is Type} block, where \texttt{Type} is the type \texttt{var} will be cast into. These type-checking statements are unlikely to be present in programs that do not already perform downcasting, thus these operators will almost always produce invalid programs.

In summary, child-parent type replacements in Dafny yield either invalid or equivalent mutants and, hence, are not useful. However, we can note that child-to-child replacement (\cref{sec:polymorphism-operators}) is a valid polymorphic mutation.

\smallskip\emph{\texttt{Static} keyword.}
MuJava defines operators that insert or delete the \texttt{static} keyword from class fields~\cite{muJava2005, muJavaClassLevelOperators}. In Java, this operator is useful because a class's behavior will differ depending on whether the updated fields are static. If a field is made static by mistake, its update in one object will cause all of the other objects of the same class to wrongfully have the same field updated to the same value. The inverse can also generate unwanted side effects.

In Dafny, only constant fields can be static. While static fields are stored in memory shared among all class instances, their constant nature means their value, and consequently program behavior, remains unchanged. Thus, these operators always generate equivalent mutants in Dafny.

\section{Novel mutation operators for Dafny}\label{chap:dafny-mutation-operators}

Mutation operators typically try to mimic errors commonly made by developers when writing a program, and it has been shown that tailored mutants are well-coupled to real faults~\cite{allamanis2016tailoredmutantsfitbugs}. These may differ based on the features of each programming language. Although most of the mutation operators presented in \Cref{chap:literature-mutation-operators} can be applied to Dafny programs, mimicking human errors in Dafny may require operators tailored for Dafny's specificities. In this section, we analyze bugfixes in real Dafny programs and synthesize them into a novel set of mutation operators for Dafny, aiming to complement the answer to \textbf{RQ1}.

\subsection{Methodology}\label{sec:commit-review-study}

To the best of our knowledge, no prior study has identified which errors commonly occur in Dafny programs. To study this, inspired by \citet{BrownOperatorMining}'s approach, we analyzed evidence of bugfixes in publicly available Dafny projects on GitHub. This analysis enabled us to extract new mutation operators tailored to Dafny, some of which are also applicable to other languages. These operators aim to simulate realistic developer errors and strengthen the relevance of our mutation testing tool, \ourTool (described in \Cref{chap:tool}).

\subsubsection{Dafny programs' repositories}

DafnyBench\footnote{\url{https://github.com/sun-wendy/DafnyBench/tree/0cd28fe}, accessed Jan. 2026.}~\cite{DafnyBench} is currently the largest dataset of Dafny programs, with 785 programs from 120 software repositories. Out of these, we noted that (a) there is no url, in DafnyBench's paper, for four repositories\footnote{BinarySearchTree, CO3408-Advanced-Software-Modelling-Assignment-2022-23-Part-2-A-Specification-Spectacular, Programmverifikation-und-synthese, and Prog-Fun-Solutions.}, (b) two were unavailable\footnote{\url{https://github.com/AoxueDing/Dafny-Projects} and \url{https://github.com/Aaryan-Patel-2001/703FinalProject}, accessed Jan. 2026.},
(c) three were repeated\footnote{\url{https://github.com/Eggy115/Dafny}, \url{https://github.com/vladstejeroiu/Dafny-programs}, and \url{https://github.com/isobelm/formal-verification}, accessed Jan. 2026.},
and (d) one was a subset of another\footnote{\url{https://github.com/secure-foundations/iron-sync} $\subset$ \url{https://github.com/secure-foundations/ironsync-osdi2023}, accessed Jan. 2026.}, resulting in the exclusion of 10 repositories. 
Additionally, we included the repositories for the AWS encryption SDK\footnote{\url{https://github.com/aws/aws-encryption-sdk/tree/1be4d62}, accessed Jan. 2026.} and Dafny-EVM\footnote{\url{https://github.com/Consensys/evm-dafny/tree/e2e52e8}, accessed Jan. 2026.}, which are recognized as key examples of Dafny programs in industry, leaving us with 112 repositories\footnote{The list of repositories can be found at \url{https://github.com/MutDafny/mutdafny-study-data/blob/main/subjects/data/generated/repositories.csv}.}.

\subsubsection{Bugfix commits}

Evidence of programming errors in software repositories may manifest in bug reports (i.e., issues labeled as \textit{bug} on GitHub) or in commit messages referencing fixes or corrections. Given that 108 of the 112 repositories had no bug reports, our analysis focused on commit messages, covering a total of 12,308 commits. To identify potential bugfixes without complete manual inspection, and aiming to achieve replicability in identifying which commit messages hint at bugfixes, we selected commit messages using the \texttt{git log} command together with \texttt{grep}, and a set of keyword patterns: 
\emph{``fix''}; \emph{``correct''}, including other variations such as \emph{``correction''}, \emph{``corrected''}; \emph{``updat''}, with variations like \emph{``update''}, \emph{``updating''}; and \emph{``chang''}, including words like \emph{``change''} and \emph{``changing''}.
This resulted in a total of 1,475 commits with messages hinting at possible bugfixes, some of which contained more than one of the listed patterns.

\subsubsection{Procedure}

Since our goal is to mutate implementations, changes involving specifications were disregarded. Each author of this paper individually analyzed one-third ($\approx$491) of the 1,475 commits\footnote{Given we aim to synthesize modifications, as mutation operators, that introduce mistakes to the code and not that fix the code, we analyzed the inverse of bugfix commits, i.e., bugfixes as bug-introducing.} and registered any changes that had the potential to be synthetized as mutation operators\,---\,e.g., modifications to expressions, altered conditions, added or removed lines of code\,---\,that could reflect common coding mistakes. We identified 270 (18.3\%) commits as potential candidates.
The authors then met to review the registered potential operators and discuss each until a consensus was reached. Of the 270, 148 were accepted, and 122 were not (\Cref{sec:non-operators} describes some reasons for this).
The list of 1,475 commits and the outcome for each can be found in \texttt{bugfixes.csv} of the \emph{supplementary material}\footnote{\label{ftn-supp-mat}\url{https://doi.org/10.6084/m9.figshare.30640202.v2}.}. The list of operators synthesized from bugfixes together with some source commits is available in the \textit{supplementary material}\footref{ftn-supp-mat}.

\subsection{Synthesized operators}\label{sec:new-dafny-operators}

Our bugfix analysis led to (i) the synthesis of 10 new mutation operators (nine that, although found in bugfixes of Dafny programs, could also be applied to programs written in other programming languages; and one tailored for Dafny), and (ii) the extension of the previously proposed mutation operator \emph{COR} to support Dafny specificities. We briefly describe the 11 operators in the following subsections. For each operator, we provide the URL to one example from the analyzed commits, and, for some, a code example of the original and mutated versions in diff format to complement their description\footnote{Each diff applies the corresponding mutation operator to an existing code, i.e., introduces a \emph{mistake}.}. Next, we complete our answer to \textbf{RQ1} by listing and describing these new mutation operators.

\subsubsection{\acf{COR}}
Dafny supports the standard conditional operators (\Cref{sec-conditional-operators}) \texttt{\&\&} and \texttt{||}, common to most traditional programming languages, as well as additional ones: \texttt{==>} (implication), \texttt{<==} (reverse implication), and \texttt{<==>} (equivalence). We found evidence\footnote{\url{https://github.com/Consensys/evm-dafny/commit/4946bec\#diff-5f537a3009d26ba61bcb2a3b39a3c4d0dbc32fff56fd177427b7e5a245dd6a3fL61-R61}, accessed Jan. 2026.} of replacements occurring between these operators in Dafny code. We have thus extended the previously defined conditional replacement operator to include them.

\subsubsection{Expression replacement}
Evidence shows expressions, in particular identifiers, to be commonly swapped.

\smallskip\emph{\acf{VER}}\footnote{\url{https://github.com/dafny-lang/Dafny-VMC/commit/27a6601\#diff-9ead0207f306a1ba28e42076fd59fe8bb6802fa2faa5288b6e43bc42d95f3b9cL28-R28}, accessed Jan. 2026.} targets variable name references. Potential replacements should be limited to in-scope variables of the same type as the original. 

\smallskip\emph{\acf{FAR}}\footnote{\url{https://github.com/secure-foundations/ironsync-osdi2023/commit/9420f92\#diff-8b6fe65b7c3a6ba702032935fb409c82b5a20dac0efcd4517966f82c1b980ee2L196-R196}, accessed Jan. 2026.}
replaces an access to a class field with another of the same class, preserving the object reference and typing, as in the following example:

\begin{lstlisting}[style=diff, numbers=none, label=lst:field-access-replacement-mutation]
class Item {
     var item: string
     var price: int
     var stock: int
}

method GetProfit(item: Item) returns (profit: int)
{
---    profit := item.price * item.stock; --
+++    profit := item.price * item.price; ++
}
\end{lstlisting}

\smallskip\emph{\acf{MCR}\footnote{\url{https://github.com/aws/aws-encryption-sdk/commit/36d1fad\#diff-d6402a5709ccca6c7b42036edacf38b136517d3b5d6206557decd0f8123522adL258-R258}, accessed Jan. 2026.} and \acf{DCR}\footnote{\url{https://github.com/secure-foundations/ironsync-osdi2023/commit/1d7eabc\#diff-dd4cf8cbf6aaf5bb44a285c7bec0890c1092315f2b6a3548b0bfa4ad4f28c2a0L666-R666}, accessed Jan. 2026.}} work similarly to \emph{\ac{VER}} and \emph{\ac{FAR}}. \emph{\ac{MCR}} replaces one method (or function) call with another: 

\begin{lstlisting}[style=diff, numbers=none, label=lst:method-call-replacement-mutation]
---    var n := Sum(10, 20); -- // return type int
+++    var n := Multiply(10, 20); ++ // return type int
\end{lstlisting}

The \emph{\ac{DCR}} operator replaces one datatype constructor call with another available from the same datatype: 

\begin{lstlisting}[style=diff, numbers=none, label=lst:datatype-constructor-replacement-mutation]
datatype MyQuantifier = None | Some(x: set<int>) | All(y: set<int>)
method Main()
{
---    var selection := Some({1, 2, 3, 4}); --
+++    var selection := All({1, 2, 3, 4}); ++
}
\end{lstlisting}

Both operators retain the original arguments. To preserve program validity, replacements are limited to callable elements with matching signatures, i.e., same return type (for \emph{MCR}) and same number, types, and order of arguments.

\smallskip\emph{\ac{MVR}}\footnote{\url{https://github.com/dafny-lang/Dafny-VMC/commit/ef3152c\#diff-66dcba1192e097932aec9462cdb8e65134fc25c93afcfa612894b2c94e64af5bL99-R99}, accessed Jan. 2026.}. Besides replacing variables with variables and method calls with other method calls, we also found evidence of method-to-variable replacements. In such cases, the replacement variable must match the return type of the target method.

\subsubsection{\acf{SAR}} A common change involves swapping the arguments of a callable element\footnote{\url{https://github.com/aws/aws-encryption-sdk/commit/b409ea5\#diff-f945451ea4d5cc43ed01a97ffb67bfabf7fc17727818427f892b90a7ce2c04e7L55}, accessed Jan. 2026.}, e.g., a method, a function, or a constructor. The corresponding mutation operator replicates this by swapping two arguments in the same call. The mutated program is only valid if the swapped arguments are of the same type.

\subsubsection{\acf{TAR}} In Dafny, tuple elements can be accessed by their positional index (as in \texttt{tuple.0})\footnote{\url{https://github.com/aws/aws-encryption-sdk/commit/4c9d5d6\#diff-00af0f655962fd28a7232ec2053eadda1257bb4587d838f9957a75c069e80437L64-R64}, accessed Jan. 2026.}. This operator replaces a tuple index with another that refers to an element of the same type.

\subsubsection{\acf{CBE}} In addition to deleting a branch of an \texttt{if} statement (see \Cref{sec:block-deletion-mutation}), our study found a common modification\footnote{\url{https://github.com/benreynwar/SiLemma/commit/4da22d6\#diff-0eff5d0d272121340e6e9d6bad6e03612a3057426e284b56cd185a1524d7937aL572-R577}, accessed Jan. 2026.} where one branch's contents are extracted to the outside of the \texttt{if} statement, and the remaining branches are deleted. Unlike the first operator, which preserves the \texttt{if} statement, this one replaces it entirely with one branch's code block, as in the following: 

\begin{lstlisting}[style=diff, numbers=none, label=lst:conditional-block-extraction-mutation2]
---    if a <= 0 then --
---        [] --
---    else --
     var b, c := a + a, a * a;
     [a, b, c]
\end{lstlisting}

\subsubsection{\acf{SWS}} A common program fix involves swapping two or more lines of code\footnote{\url{https://github.com/dafny-lang/libraries/commit/5ecd461\#diff-3a16ac84e5bda345c586253ae5dbaa34730085d17a587393650195f5248eea80L46-R48}, accessed Jan. 2026.}, affecting logical flow and program behavior. While the concept is too broad to be fully translated into a mutation operator, due to the exponential mutant possibilities that would result from swapping every group of lines in a program, simpler operations can approximate it. We propose swapping a statement with the one directly above or below it.

\subsubsection{\acf{SWV}} Lastly, the \emph{\ac{SWV}} operator reflects another case related to the swapping of program constructs, this time, applied to the declaration of variables in the same scope\footnote{\url{https://github.com/Consensys/evm-dafny/commit/58abe25\#diff-eec5c501c95e783482e4e062ad3c9c30e947c394e0388bc197ad2306a5840ea5L315-R316}, accessed Jan. 2026.}. 

\begin{lstlisting}[style=diff, numbers=none, label=lst:var-decl-swap-mutation]
---    var perimeter := 2.0 * radius * 3.14; --
---    var area := radius * radius * 3.14; --
+++    var perimeter := radius * radius * 3.14; ++
+++    var area := 2.0 * radius * 3.14; ++
\end{lstlisting}

This operator is similar to \emph{\ac{VER}}, but it affects the entire scope of the variable, while \emph{\ac{VER}} only affects the single place where the variable is replaced.

\subsection{Excluded modifications}\label{sec:non-operators} 

Many commits contained modifications that, at first glance, seemed promising, especially to those unfamiliar with Dafny's syntax, but that were ultimately not included as mutation operators. We can highlight three groups of common modifications: (a) changes that alone do not alter program behavior, (b) changes inherent to migrations from Dafny 3 to Dafny 4, and (c) code improvements.

The first group includes, e.g., replacing types that have subset relationships (like \texttt{nat} with \texttt{int}) and type parameter property changes\footnote{\url{https://github.com/secure-foundations/ironsync-osdi2023/commit/e27b26c\#diff-7091ca089a282c06587ac4406db89d238ea977e5e47b93c290d769c1221f2be3L14-R14}, accessed Jan. 2026.}\textsuperscript{,}\footnote{\url{https://dafny.org/dafny/DafnyRef/DafnyRef.html\#sec-type-characteristics}, accessed Jan. 2026.}.
The second relates to syntax changes introduced by Dafny 4 in how functions and callables are declared\footnote{\url{https://github.com/namin/dafny-sandbox/commit/21e148e\#diff-4698b1118ee26951776954c2d28642a7dc6ad3abb38743d0034655ffdf302364L112-R112}, accessed Jan. 2026.}. For example, functions were \texttt{ghost} by default (i.e., not compiled, only available for verification purposes), but now all functions compile unless declared as \texttt{ghost}. Finally, the last group includes common code improvement changes, e.g., the replacement of the declaration of a callable element as a method, function, predicate, or lemma, and the insertion or deletion of \texttt{ghost} keywords\footnote{\url{https://github.com/secure-foundations/ironsync-osdi2023/commit/0da7158}, accessed Jan. 2026.}\textsuperscript{,}\footnote{\url{https://github.com/secure-foundations/ironsync-osdi2023/commit/de87232\#diff-10aacdf18deb66491a2e4bb876eaeff2e7979b41d427563229817337b5302753L412-R420}, accessed Jan. 2026.}, which optimizes compilation by eliminating the need to compile certain elements, without affecting program behavior.

\begin{figure*}[tb]
  \centering
  \includegraphics[width=\textwidth]{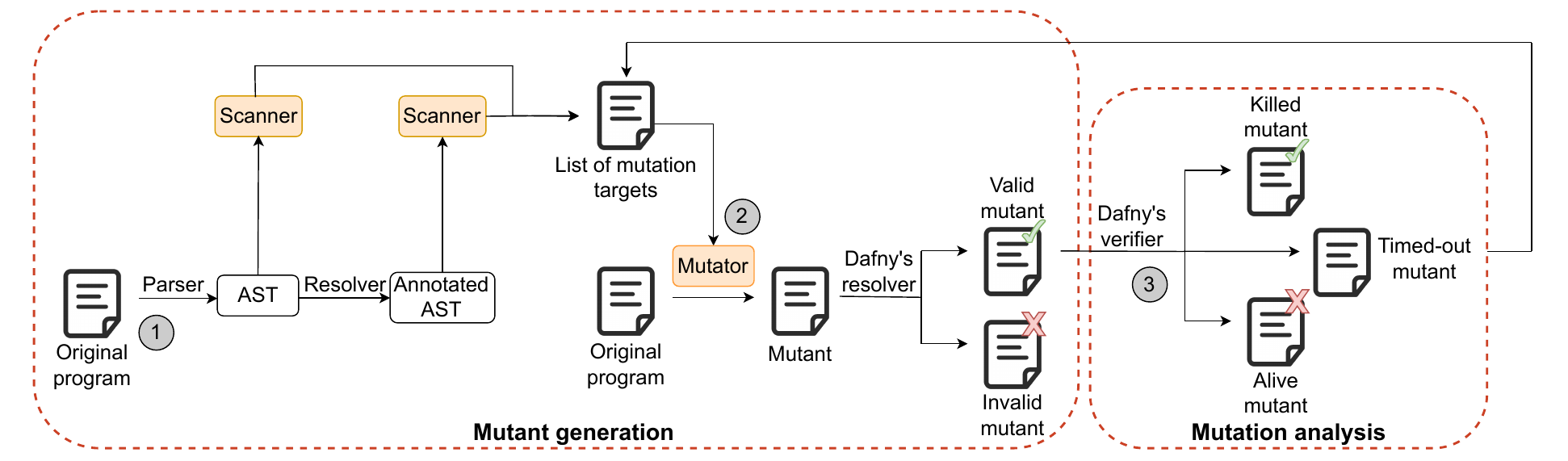}
  \caption{Mutation testing pipeline with \ourTool. \ourTool's components are depicted by the filled boxes, the rest corresponding to Dafny.\label{fig:full-solution}}
\end{figure*}

\section{\ourTool}\label{chap:tool}

\ourTool supports 40 mutation operators\footnote{The full list of operators implemented in \ourTool can be consulted in the \textit{supplementary material} at \url{https://doi.org/10.6084/m9.figshare.30640202.v2}.}: the 30 described in \Cref{chap:literature-mutation-operators} and the 10 described in \Cref{chap:dafny-mutation-operators}. The tool is integrated into the Dafny verification framework, implemented as a plugin. Dafny's plugin architecture provides an extensible interface for \ac{AST} analysis and manipulation, allowing \ourTool to integrate directly with Dafny's execution flow without modifying its source code. \ourTool works at the level between the parser and verifier and consists of a two-phase pipeline: \emph{mutant generation} and \emph{mutation analysis}, illustrated in \Cref{fig:full-solution}.

\smallskip\textbf{Mutant generation.}
Operates by directly manipulating Dafny's internal representation. A \emph{scanner} component traverses the \ac{AST} to identify \emph{mutation targets}\,---\,specific locations in the program where specific types of mutations can be applied\,---\,and collects the necessary context for performing those mutations. Namely, for each target, \ourTool records the node's position, the type of mutation operator to be used, and any additional arguments required to apply the mutation.
The \emph{mutator} component processes each of the identified mutation targets and changes the program's AST to create individual mutants prior to Dafny's resolution phase. This is done by traversing the \ac{AST} until the target's location, i.e., the node to mutate, is reached, and then performing the necessary operations on it. The changes can involve altering a node's data, replacing an expression or statement with a different one by creating a new node (e.g., the \emph{\ac{BBR}} operator involves the replacement of a binary expression with a boolean literal one), or deleting nodes by removing their reference from the parent.

For certain mutation operators, identifying mutation targets involves analyzing the type information of the target node, which is only available post-resolution. An example is the \emph{AOI} operator, which can only be applied to numerical expressions. Here, to know whether a tree node constitutes a valid mutation target, we must first know its type. As a result, and as depicted in \Cref{fig:full-solution}, a portion of the targets are identified during a pre-resolve tree visit, and the remaining are identified during a post-resolve one.

\smallskip\textbf{Mutation analysis.}
The generated mutants are evaluated using Dafny's verification capabilities and the specification under test. Mutants are classified according to the verification outcome\,---\,\emph{Alive (survived):} the mutant verifies successfully against the original spec, hinting at a potential specification weakness; \emph{Killed:} the mutant fails verification, meaning that the spec successfully detected the inserted fault; \emph{Invalid:} the mutant fails during Dafny's resolution process (before verification), indicating a semantically incorrect program; \emph{Timed Out:} the verifier could not determine an outcome within a set time limit\footnote{By default, Dafny's verifier runs for 20 seconds.}.

\section{Empirical Evaluation}\label{chap:eval}

In this section, we describe and discuss the results of our empirical evaluation, which aims to answer the following research questions:

\begin{itemize}[label={}, leftmargin=0pt, itemindent=0pt]
  \item \textbf{RQ2:} Are mutation operators for Dafny programs helpful in the identification of specification weaknesses?
  \item \textbf{RQ3:} How effective are the different mutation operators?
  \item \textbf{RQ4:} How efficient is \ourTool at generating mutants?
\end{itemize}

\subsection{Experimental procedure}\label{chap:eval:sec:procedure}

The empirical evaluation of \ourTool was conducted on a dataset with 43,178 lines of code consisting of 743 Dafny files\footnote{In this paper, we consider a Dafny \emph{file}, \text{.dfy}, as a Dafny \emph{program} (as defined in DafnyBench~\cite{DafnyBench}), and we may use the terms \emph{file} or \emph{program} interchangeably, in which case we imply the former.} from DafnyBench~\cite{DafnyBench}, 17 from AWS encryption SDK, and 34 from Dafny-EVM.  
DafnyBench includes 785 files, but 42 of these were unverifiable with our environment. 
We excluded any files that (a) had validity errors with our Dafny/Z3 version, (b) timed out, (c) had verification errors, and (d) verified nothing at all (e.g., \texttt{Dafny program verifier finished with 0 verified, 0 errors})\footnote{Examples of excluded files: (a) \texttt{groupTheory\_tmp\_tmppmmxvu8h\_assign\\ment1}; (b) \texttt{WrappedEther}; (c) \texttt{Clover\_insert}; and (d) \texttt{DafnyExercises\_\\tmp\_tmpd6qyevja\_QuickExercises\_testing2}. The full list of files and the reason why each did not verify can be found in the \emph{supplementary material} at \url{https://doi.org/10.6084/m9.figshare.30640202.v2}.}.

The 794 program files that constitute our experimental subjects gather a total of 2,747 methods, functions, and similar constructs: 1,668 of these have no pre-conditions, 1,403 no post-conditions, and 1,044 have neither. This does not necessarily imply that the spec is \emph{weak}. In fact, it is common for functions and predicates to omit such contracts. However, when these are invoked within methods that \emph{do} have specs, mutations affecting them can still contribute to measuring the quality of the overall program's specification.

Our experimental procedure consists of running \ourTool against the 794 files.
To accelerate the experiment runtime, we parallelized some workflow components. Each scanner job involves scanning a single program for mutation targets using a single mutation operator, while for the mutator, each job entails generating mutants for a single program. We executed each experiment (i.e., a mutation operator on a file) 10 times. We then removed 20\% of the repetitions with the highest and the lowest runtimes.

The experiment was run on a cluster running Ubuntu 20.04, Linux kernel version 5.15, with 1GB of RAM, and an Intel(R) E5-2450 CPU at 2.1GHz. We use Dafny 4.10.0. (version corresponding to commit \texttt{7159879}\footnote{\url{https://github.com/dafny-lang/dafny/tree/7159879}, accessed Jan. 2026.}) and Z3 v4.12.6.

\subsection{Experimental metrics}\label{chap:eval:sec:metrics}

The dataset resulting from running this experiment includes all generated mutants, runtime measurements and mutation target data for all file scans, runtime measurements for every mutant generation, and data identifying the applied mutation and its analysis status. For each execution, we record the total runtime of each scan/\allowbreak mutation process and the time spent in each of Dafny's workflow stages and executing the plugin. This data allows us to compute the mutant generation, mutation analysis, and total runtime.

Additionally, we compute the \emph{mutation score}~\cite{5487526}, i.e., the ratio of killed mutants to the total number of mutants, excluding invalid mutants, e.g., that introduce non-compiling changes, and time-out ones, which reflect an inconclusive result. This metric reflects the fault-detection capability, typically of the test suite, and, in the context of this work, of the specifications. A higher mutation score increases confidence in the test suite/specification.

\subsection{Experimental results}\label{chap:eval:sec:results}

Overall, \textbf{\mbox{\ourTool} generated a total of 118,458 mutants} across our dataset of 794 programs, an average of 2.7 mutants per line of code.

\subsubsection{\textbf{Answer to RQ2}}

\begin{figure}[tb]
  \centering
  \includegraphics[width=\columnwidth]{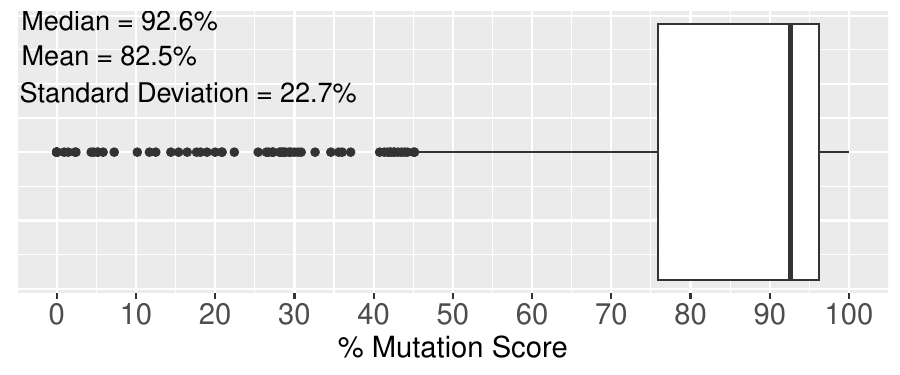}
  \vspace{-2em}\caption{Mutation score distribution.\label{fig:mutation-score}}
\end{figure}

We compute the mutation score of each program in the dataset, illustrated in \mbox{\Cref{fig:mutation-score}}. On average, the dataset's \textbf{specifications are able to kill 82\% of the mutants}. The middle 50\% scores range from 76\% to 96\% with a standard deviation of 23\%, indicating a concentration of high mutation scores. These results show that Dafny specifications exhibit high mutant-detection capability. Still, a single alive mutant may be enough to reveal a weakness.

From the 30,459 surviving mutants, 11,035 affect methods with post-conditions. The remaining 19,424 can be weak due to a lack of specification in the mutated component, hint at a weakness in a component calling the mutated one, or be equivalent to the original program. Due to the infeasibility of manually analyzing 11,035 mutants to assess whether they survived due to a weakness or equivalence to the original program, we focused on 24 randomly selected files and their 284 surviving mutants in methods with post-conditions. This took eight hours of one person's effort ($\approx$ 1.7 minutes per mutant), with the files averaging 50.2 lines of code (1,205 total) and 3.3 functions (79 total).

In these 24 files, we found five specification weaknesses\footnote{\label{ftn-deleted-repo}\url{https://github.com/sun-wendy/DafnyBench/blob/main/DafnyBench/dataset/ground_truth/BinarySearchTree_tmp_tmp_bn2twp5_bst4copy.dfy\#L58}, accessed Jan. 2026.}\textsuperscript{,}\footnote{\label{ftn-dafny-synthesis-report}\url{https://github.com/sun-wendy/DafnyBench/blob/main/DafnyBench/dataset/ground_truth/dafny-synthesis_task_id_126.dfy\#L1}, accessed Jan. 2026.}\textsuperscript{,}\footnote{\label{ftn-prev-reported}\url{https://github.com/sun-wendy/DafnyBench/blob/main/DafnyBench/dataset/ground_truth/dafny-synthesis_task_id_161.dfy\#L7}, accessed Jan. 2026.}\textsuperscript{,}\footnote{\label{ftn-motivational-ex}\url{https://github.com/sun-wendy/DafnyBench/blob/main/DafnyBench/dataset/ground_truth/dafny-synthesis_task_id_2.dfy\#L7}, accessed Jan. 2026.}\textsuperscript{,}\footnote{\label{ftn-new-report}\url{https://github.com/sun-wendy/DafnyBench/blob/main/DafnyBench/dataset/ground_truth/dafny-synthesis_task_id_249.dfy\#L7}, accessed Jan. 2026.} in five programs from two different projects, one\footref{ftn-motivational-ex} corresponding to the motivational example presented in \Cref{chap:intro}. The repository of one of the weak specs\footref{ftn-deleted-repo} became unavailable, so we could not contact its developers. The remaining programs were generated by LLMs for a study~\cite{DafnySynthesis}, and a manual inspection led \citet{DafnySynthesis} to report one of the weaknesses\footref{ftn-dafny-synthesis-report} (confirming our finding). We contacted them for reporting the remaining three that went unnoticed\footnote{\url{https://github.com/Mondego/dafny-synthesis/issues/2}, accessed Jan. 2026.} and, while Misu et al.\ acknowledged our discoveries, it was also pointed out that two of the weaknesses\footref{ftn-prev-reported}\textsuperscript{,}\footref{ftn-motivational-ex} had previously been detected by \citet{Endres2024}, the third one\footref{ftn-new-report} being a new discovery.

In the case of \texttt{bst4copy}\footref{ftn-deleted-repo}, the \texttt{insert} method ensures that the output tree is a valid binary search tree, but there is no condition asserting that the insertion of the input value indeed occurred. Similarly, the \texttt{insertRecursion} method checks the relation between the tree's lowest and highest value and the value to be inserted, but does not check if the latter is present in the output tree. If no modification is made to the input tree, neither method's specification would be able to detect the incorrect behavior. In \texttt{task\_id\_126}\footref{ftn-dafny-synthesis-report}, the post-condition states that every divisor common to two input numbers is lower than their sum. However, if an incorrect implementation caused only some of the common divisors to be added to the sum, this single post-condition would not be able to detect it. The purpose of \texttt{task\_id\_161}\footref{ftn-prev-reported} is to return every unique element present in the first input array but not in the second. While the specification states that every element in the result must be in the first array but not in the second, it does not enforce that every element in the first array that is not present in the second must be a part of the result, making empty lists a valid output. \texttt{task\_id\_2}\footref{ftn-motivational-ex} and \texttt{task\_id\_249}\footref{ftn-new-report} are very similar to \texttt{task\_id\_161}, the goal being to compute the unique elements common to two input arrays, and the weakness being the same. All of these weaknesses fail to address a crucial guarantee.

Of the 284 analyzed mutants, 77 hinted at these weaknesses and were successfully killed upon strengthening the specifications, 157 were equivalent to the original program, and we could not draw meaningful conclusions from the remaining 50. The number of equivalent mutants, along with the time to perform manual verification of the surviving mutants, poses an evident challenge. In practice, we expect \ourTool to be used as a validation tool alongside development to incrementally validate new components rather than entire projects at once. This will result in fewer mutants per iteration and shorten analysis time, making it feasible.

\subsubsection{\textbf{Answer to RQ3}}

{\begin{table}[tb]
\caption{Number of generated, killed, survived, invalid, and timeout mutants.\label{tab:mutants-status}}
\vspace{-1em}\resizebox{\columnwidth}{!}{
\begin{tabular}{@{}lrrrrr@{}} \toprule
\textbf{Op.} & \textbf{\# Mut} & \textbf{\# Killed} & \textbf{\# Survived} & \textbf{\# Invalid} & \textbf{\# Timeout} \\ \midrule
AMR & 0 & 0 (0.0\%) & 0 (0.0\%) & 0 (0.0\%) & 0 (0.0\%) \\
AOD & 250 & 199 (79.6\%) & 51 (20.4\%) & 0 (0.0\%) & 0 (0.0\%) \\
AOI & 15,787 & 13,650 (86.5\%) & 1,940 (12.3\%) & 122 (0.8\%) & 75 (0.5\%) \\
AOR & 6,202 & 4,963 (80.0\%) & 794 (12.8\%) & 371 (6.0\%) & 74 (1.2\%) \\
BBR & 5,722 & 3,695 (64.6\%) & 1,869 (32.7\%) & 144 (2.5\%) & 14 (0.2\%) \\
CBR & 369 & 3 (0.8\%) & 364 (98.6\%) & 2 (0.5\%) & 0 (0.0\%) \\
CIR & 926 & 657 (71.0\%) & 233 (25.2\%) & 36 (3.9\%) & 0 (0.0\%) \\
COD & 76 & 64 (84.2\%) & 12 (15.8\%) & 0 (0.0\%) & 0 (0.0\%) \\
COI & 3,851 & 2,883 (74.9\%) & 814 (21.1\%) & 142 (3.7\%) & 12 (0.3\%) \\
COR & 1,792 & 839 (46.8\%) & 731 (40.8\%) & 216 (12.1\%) & 6 (0.3\%) \\
EVR & 13,976 & 10,080 (72.1\%) & 3,276 (23.4\%) & 557 (4.0\%) & 63 (0.5\%) \\
LBI & 738 & 621 (84.2\%) & 115 (15.6\%) & 1 (0.1\%) & 1 (0.1\%) \\
LOD & 1 & 0 (0.0\%) & 1 (100.0\%) & 0 (0.0\%) & 0 (0.0\%) \\
LOI & 33 & 19 (57.6\%) & 11 (33.3\%) & 2 (6.1\%) & 1 (3.0\%) \\
LOR & 20 & 9 (45.0\%) & 10 (50.0\%) & 0 (0.0\%) & 1 (5.0\%) \\
LSR & 54 & 30 (55.6\%) & 24 (44.4\%) & 0 (0.0\%) & 0 (0.0\%) \\
LVR & 13,927 & 9,860 (70.8\%) & 4,005 (28.8\%) & 31 (0.2\%) & 31 (0.2\%) \\
MAP & 5,399 & 345 (6.4\%) & 2,512 (46.5\%) & 2,539 (47.0\%) & 3 (0.1\%) \\
MMR & 20 & 10 (50.0\%) & 10 (50.0\%) & 0 (0.0\%) & 0 (0.0\%) \\
MNR & 421 & 124 (29.4\%) & 237 (56.3\%) & 60 (14.2\%) & 0 (0.0\%) \\
MRR & 2,286 & 976 (42.7\%) & 433 (18.9\%) & 874 (38.2\%) & 3 (0.1\%) \\
ODL & 4,418 & 2,115 (47.9\%) & 265 (6.0\%) & 2,014 (45.6\%) & 24 (0.5\%) \\
PRV & 0 & 0 (0.0\%) & 0 (0.0\%) & 0 (0.0\%) & 0 (0.0\%) \\
ROR & 13,180 & 8,187 (62.1\%) & 3,996 (30.3\%) & 978 (7.4\%) & 19 (0.1\%) \\
SDL & 5,952 & 3,269 (54.9\%) & 1,301 (21.9\%) & 1,369 (23.0\%) & 13 (0.2\%) \\
SLD & 127 & 85 (66.9\%) & 41 (32.3\%) & 0 (0.0\%) & 1 (0.8\%) \\
SOR & 2 & 2 (100.0\%) & 0 (0.0\%) & 0 (0.0\%) & 0 (0.0\%) \\
THI & 8 & 5 (62.5\%) & 1 (12.5\%) & 2 (25.0\%) & 0 (0.0\%) \\
THD & 0 & 0 (0.0\%) & 0 (0.0\%) & 0 (0.0\%) & 0 (0.0\%) \\
VDL & 1,248 & 196 (15.7\%) & 589 (47.2\%) & 460 (36.9\%) & 3 (0.2\%) \\
\midrule
\textit{Total} & 96,785 & 62,886 (65.0\%) & 23,635 (24.4\%) & 9,920 (10.2\%) & 344 (0.4\%) \\
\midrule
CBE & 1,801 & 1,217 (67.6\%) & 564 (31.3\%) & 15 (0.8\%) & 5 (0.3\%) \\
DCR & 415 & 243 (58.5\%) & 113 (27.2\%) & 57 (13.7\%) & 2 (0.5\%) \\
FAR & 93 & 60 (64.5\%) & 30 (32.3\%) & 3 (3.2\%) & 0 (0.0\%) \\
MCR & 478 & 112 (23.4\%) & 349 (73.0\%) & 17 (3.6\%) & 0 (0.0\%) \\
MVR & 1,972 & 1,258 (63.8\%) & 481 (24.4\%) & 226 (11.5\%) & 7 (0.3\%) \\
SAR & 923 & 458 (49.6\%) & 401 (43.5\%) & 60 (6.5\%) & 4 (0.4\%) \\
SWS & 3,671 & 1,258 (34.3\%) & 2,194 (59.8\%) & 206 (5.6\%) & 13 (0. \%) \\
SWV & 400 & 188 (47.0\%) & 153 (38.2\%) & 59 (14.8\%) & 0 (0.0\%) \\
TAR & 16 & 8 (50.0\%) & 3 (18.8\%) & 5 (31.2\%) & 0 (0.0\%) \\
VER & 11,904 & 8,829 (74.2\%) & 2,536 (21.3\%) & 428 (3.6\%) & 111 (0.9\%) \\
\midrule
\textit{Total} & 21,673 & 13,631 (62.9\%) & 6,824 (31.5\%) & 1,076 (5.0\%) & 142 (0.7\%) \\
\midrule
\textit{Overall Total} & 118,458 & 76,517 (64.6\%) & 30,459 (25.7\%) & 10,996 (9.3\%) & 486 (0.4\%) \\
\bottomrule
\end{tabular}}
\end{table}}

\Cref{tab:mutants-status} reports, for every mutation operator, how many of the mutants generated with it were killed, survived, invalid, or whose verification timed out. \emph{AOI}, \emph{ROR}, \emph{EVR}, \emph{LVR}, and \emph{VER} are responsible for the generation of the higher number of mutants, which is expected since they target common syntactic constructs: respectively, binary operators, expressions, and variables. \emph{AMR}, \emph{PRV}, and \emph{THD} do not generate any mutants, and \emph{THI}, \emph{TAR}, \emph{SOR}, and \emph{LOD} generate a low number (less than 20 each). This happens because each of these operators can only be applied in very specific scenarios. For example, \emph{THD} entails a program location where a class field is accessed using the \texttt{this} keyword inside a method whose argument has the same name as that field. \emph{THI} entails the opposite. Furthermore, logical unary and shift operators occur very rarely in our dataset.

{\begin{table}[tb]
\caption{The number of mutants generated by each mutation operator that led to the identification of weak specifications.}\label{tab:weakness-ops}
\vspace{-1em}\resizebox{\columnwidth}{!}{
\fontsize{20}{24}\selectfont
\begin{tabular}{@{}lrrrrrr@{}} \toprule
\textbf{Op.} & \texttt{bst4copy\footref{ftn-deleted-repo}} & \texttt{task\_id\_126\footref{ftn-dafny-synthesis-report}} & \texttt{task\_id\_161\footref{ftn-prev-reported}} & \texttt{task\_id\_2\footref{ftn-motivational-ex}} & \texttt{task\_id\_249\footref{ftn-new-report}} & \textbf{Total} \\ \midrule
EVR & 1 & 5 & 4 & 1 & 4 & 15 \\
VER & 2 & 11 & 1 & 0 & 0 & 14 \\
ROR & 5 & 2 & 0 & 0 & 0 & 7 \\
BBR & 2 & 3 & 1 & 0 & 1 & 7 \\
COR & 0 & 4 & 1 & 0 & 1 & 6 \\
MAP & 6 & 0 & 0 & 0 & 0 & 6 \\
MVR & 4 & 0 & 0 & 0 & 0 & 4 \\
SDL & 0 & 0 & 2 & 0 & 2 & 4 \\
DCR & 3 & 0 & 0 & 0 & 0 & 3 \\
LBI & 0 & 0 & 1 & 1 & 1 & 3 \\
CBE & 1 & 1 & 0 & 0 & 0 & 2 \\
CIR & 0 & 0 & 1 & 0 & 1 & 2 \\
SWS & 0 & 0 & 1 & 0 & 1 & 2 \\
LVR & 0 & 1 & 0 & 0 & 0 & 1 \\
UOI & 1 & 0 & 0 & 0 & 0 & 1 \\
\midrule
\textit{Total} & 25 & 27 & 12 & 2 & 11 & 77 \\
\bottomrule
 \end{tabular}}
\end{table}}

As summarized in \Cref{tab:weakness-ops}, the five specification weaknesses unveiled during our manual analysis resulted from mutants generated by 10 of the existing and five of the newly proposed operators. \emph{EVR}, \emph{VER} (new), and \emph{ROR} were the ones that generated the largest number of weakness-revealing mutants. These do not correspond to the operators that, out of the 40, generate the highest percentages of survivors; only around 20-30\%. Instead, they generate more mutants overall, which suggests a correlation between a high mutant count and a greater likelihood of discovering weaknesses.

Regarding the relative effectiveness between the previously proposed and the novel mutation operators, the Wilcoxon test (with a confidence level of 99\%) reports that programs' specifications are statistically significantly (\emph{p}-value $= 1.749 \times 10^{11}$) more effective at killing mutants generated by the former than by the latter. These results further motivate one of our contributions, highlighting the need for mutation operators tailored for Dafny to more effectively evaluate Dafny specifications.

\subsubsection{\textbf{Answer to RQ4}}

\begin{figure}[tb]
  \centering
  \includegraphics[width=\columnwidth]{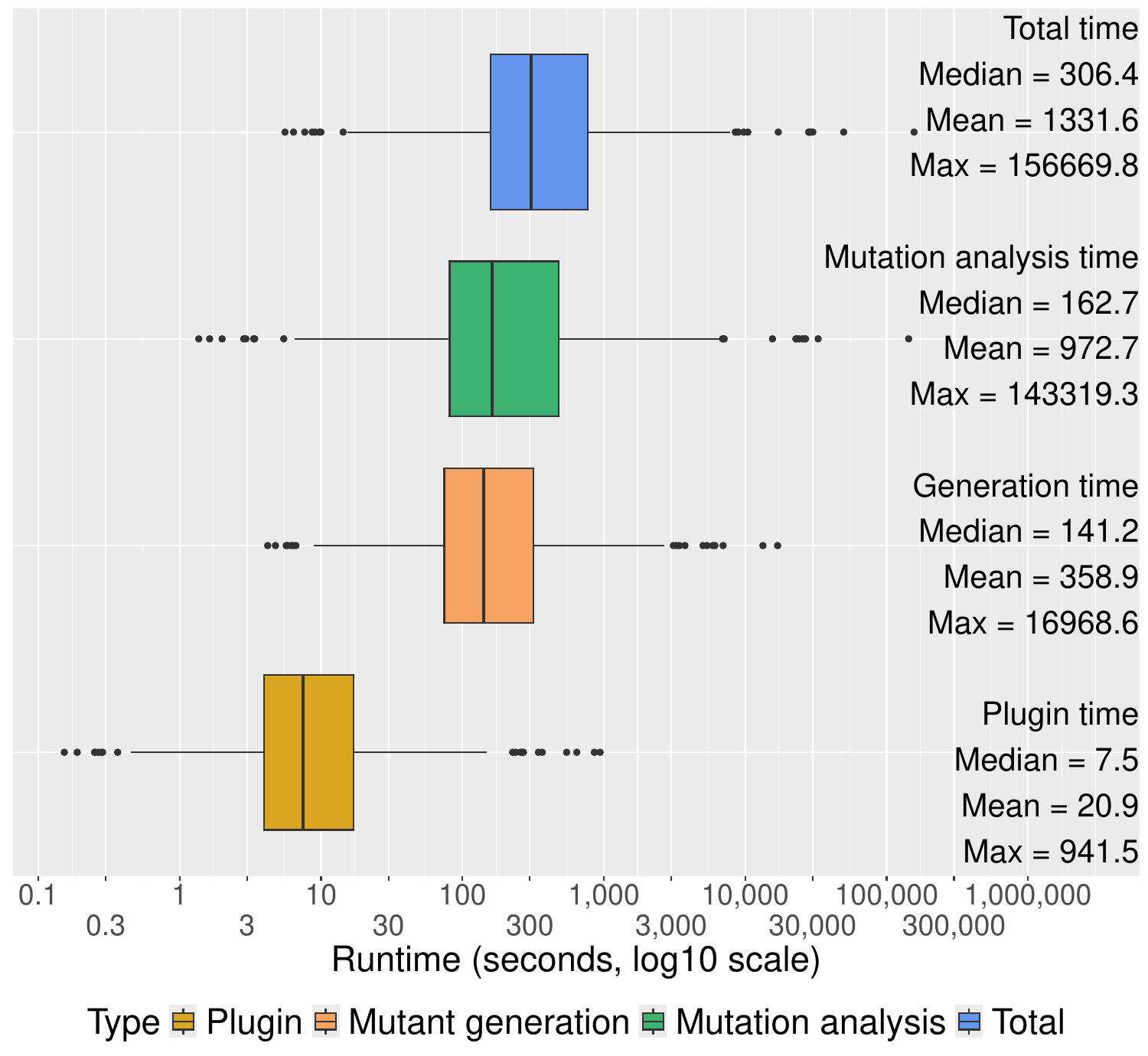}
  \vspace{-2em}\caption{Plugin, mutant generation, mutation analysis, and total runtime to perform mutation testing of a Dafny file.\label{fig:overall-runtime}}
\end{figure}

\Cref{fig:overall-runtime} shows the distribution of the runtime of the mutation testing process of each program. \textbf{The average total runtime is 22 minutes, with half of the experiments falling between two and 13 minutes.} These runtimes may be acceptable for the analysis of individual files, but can raise scalability concerns for multi-file projects. However, these issues can be mitigated through parallelization, as done in our experimental procedure.

The two phases of the mutation testing pipeline\,---\,generation and analysis\,---\,present similar average runtimes. Therefore, we conclude that the total runtime is approximately divided between mutant generation and verification.

While \ourTool usually takes one to five minutes to generate all the mutants for one file, this does not accurately reflect our implementation's runtime. In half the files, the plugin scans and mutates the program in four to 17 seconds, accounting for about 5\% of the mutant generation runtime. The remaining time is spent on mutant resolution, a regular part of Dafny's execution but crucial for identifying invalid mutants. \ourTool's efficiency is, thus, mainly affected by the resolution and verification of the mutated file.

\subsection{Threats to validity}\label{chap:eval:sec:threats}
In this subsection, we discuss threats to validity according to the guidelines defined by \citet{wohlin2012experimentation} and \citet{yin2009case}.

\smallskip\textbf{Threats to construct validity.}
We identify two main constructs in our study: \ourTool's integration with Dafny and specification effectiveness. Regarding the former, we define how the tool integrates into Dafny's workflow, identifying the mutation and verification points. For the latter, we adopt the standard mutation score to evaluate specification effectiveness, clearly defining mutant statuses and detection criteria to align with established mutation measurements. %

\smallskip\textbf{Threats to internal validity.}
The main threats we identify are CPU warm-up distortion and software bugs in \ourTool. 
The first was mitigated by running the experiment 10 times and removing outliers. Regarding software bugs, we employed testing and manually validated a subset of results. Our source code and experimental results are publicly available to support verification and reproducibility.

\smallskip\textbf{Threats to external validity.}
To support external validity, we rely on the representativeness of DafnyBench, which includes synthesized and real-world program files, the latter constituting 75\% of the dataset and having been collected from GitHub with minimal filtering. Programs vary in complexity, some comprising introductory examples written by new learners of Dafny, thus with less relevance for rigorous assessment. Yet, all include specifications, allowing the evaluation of their response to the tool. In addition, the files from AWS and Dafny-EVM are recognized as key industry examples. Nevertheless, the limited availability of high-quality open-source Dafny repositories is still a challenge.

\section{Future Work}\label{chap:future-work}

It is crucial to ease the manual examination of surviving mutants by minimizing the number of weakness hints that are either equivalent or false positives. As future work, we aim to investigate the application of code coverage of Dafny specifications to minimize the number of generated mutants~\cite{major2011}, i.e., if a particular code portion is not covered by any existing specification in the program, then no mutant in it can ever be killed. In addition, we aim to explore automatic techniques to find and discard equivalent~\cite{Papadakis2015,Houshmand2017,Medusa2019,Papadakis2018,KushigianEquivMut2024} or ineffective~\cite{8240964} mutants in the context of Dafny.

\section{Conclusion}\label{chap:concl}

We presented a methodology for verification-aware languages and its implementation in a novel tool for detecting specification weaknesses through mutation of Dafny programs, benefiting from direct integration with the language's workflow. Another key contribution of this paper is the systematic study of mutation operators suitable for Dafny, resulting in the synthesis of 30 unique operators previously proposed for other programming languages, as well as 10 newly synthesized from evidence collected in real-world Dafny bugfixes.

Our empirical evaluation across 794 Dafny programs and 118,458 mutants demonstrated the value of \ourTool in detecting weaknesses. The results show that Dafny specifications have a high mutant-detection capability. However, the analysis of a sample of alive mutants revealed five weaknesses in the benchmark programs. Despite the runtimes being highly conditioned by Dafny's verification pipeline and the lengthy manual inspection of alive mutants, the tool can be feasibly applied in real-world settings. Ultimately, \ourTool has the potential to advance formal verification by aiding Dafny developers in improving system confidence.

\begin{acks}
Isabel Amaral was financed by National Funds through the FCT - Fundação para a Ciência e a Tecnologia, I.P.\ (Portuguese Foundation for Science and Technology) within the project VeriFixer, with reference 2023.15557.PEX (DOI: \href{https://doi.org/10.54499/2023.15557.PEX}{10.54499/2023.15557.PEX}). Alexandra Mendes is funded by national funds through FCT – Fundação para a Ciência e a Tecnologia, I.P., under the support UID/50014/2025 (\href{https://doi.org/10.54499/UID/50014/2025}{https://doi.org/10.54499/UID/50014/2025}). José Campos was supported by the LASIGE Research Unit, ref.\ UID/00408/2025, DOI \href{https://doi.org/10.54499/UID/00408/2025}{10.54499/UID/00408/2025}. %
\end{acks}

\bibliographystyle{ACM-Reference-Format}
\bibliography{main}

\end{document}